\documentstyle[aps,floats,epsf,psfig,twocolumn]{revtex}
\tighten
\begin{document}
\draft 
\title{Extended Impurity Potential in a ${\mbox{\boldmath$d_{x^2-y^2}$}}$ 
Superconductor}
\author{A.P. Kampf$^{1}$ and T.P. Devereaux$^{2}$}
\address{$^{1}$
Institut f\"ur Theoretische Physik, Universit\"at zu K\"oln,\\
Z\"ulpicher Str. 77, 50937 K\"oln, Germany}
\address{$^{2}$
Department of Physics, George Washington University,\\
Washington, D.C. 20052}
\address{~
\parbox{14cm}{\rm 
\medskip
We investigate the role of a finite potential range of a nonmagnetic impurity 
for the local density of states in a $d_{x^2-y^2}$ superconductor. Impurity 
induced subgap resonances are modified by the appearance of further scattering 
channels beyond the $s$--wave scattering limit. The structure of the local 
density of states (DOS) in the vicinity of the impurity is significantly 
enhanced and therefore improves the possibility for observing the 
characteristic anisotropic spatial modulation of the local DOS in a 
$d_{x^2-y^2}$ superconductor by scanning tunneling microscopy.  
}}
\maketitle

\narrowtext
A lot of attention has been focussed recently on the role of impurities in 
unconventional superconductors. The reason for this interest is that impurities
modify the superconducting properties in a way which is characteristic for the
pairing state and thereby serve as a diagnostic tool for its identification. In
particular, magnetic impurities act as strong pair breakers for superconductors
in a spin singlet pairing state while already non--magnetic impurities are pair
breakers in superconductors with a nontrivial phase of the pairing amplitude. 
The latter kind of defects or impurities have only little effect on the 
transition temperature and the superfluid density in conventional $s$--wave 
superconductors as understood from Anderson's theorem \cite{Anderson}. 

Regarding high--$T_c$ superconducting materials evidence has accumulated for an
anisotropic energy gap most likely of $d_{x^2-y^2}$ symmetry \cite{Ginsberg}. 
For this pairing state nonmagnetic impurities produce a finite lifetime for 
quasiparticles near the nodes of the gap and a finite DOS at low energies 
\cite{Goldenfeld,Peter,Sascha}. Indeed, the measured low temperature properties
of high--$T_c$ materials show a remarkable sensitivity for impurity effects. 
Examples are the $T^2$ variation of the low temperature penetration depth 
\cite{Hardy} and the Knight shift in $Zn$ doped $YBCO$ \cite{Ishida}, and the
$\omega^3$ to $\omega$ crossover in the low frequency $B_{1g}$ Raman intensity 
\cite{Devereaux}.    

A general feature of pair breaking impurities in $s$--wave superconductors is 
that they lead to bound quasiparticle states in the energy gap \cite{Yu}. 
Similarly, for a $d$--wave superconductor with a particle--hole continuum and a
DOS extending linearly to zero frequency, low energy resonances are created 
with a highly anisotropic structure \cite{Byers,Choi,Balatsky}. Increasing the 
impurity concentration these subgap resonances form bands which give rise to 
the afore mentioned finite DOS at zero energy. The local structure of the 
resonant states in the vicinity of the impurity is characteristic of the 
pairing state of the superconductor. It has therefore been suggested that 
probing these impurity states by scanning tunneling microscopy (STM) offers a 
direct way for measuring the anisotropic structure of the pairing state 
\cite{Salkola}. However, the predicted effect may be too small to detect with 
the current STM spatial resolution and thus any advances that could enhance a 
characteristic signal would be helpful towards making STM a feasable probe to 
detect the gap symmetry \cite{Maggio}.

Systematic studies in high--$T_c$ materials have been performed by substituting
$Zn$ on the planar $Cu$ sites and thereby suppressing the local moment. Model 
calculations, as initiated by the early work of Annett et al. and Hirschfeld 
and Goldenfeld \cite{Goldenfeld}, which assume a $d_{x^2-y^2}$ gap function 
with unitary $s$--wave impurity scattering have provided a consistent 
explanation for the experimentally observed transport properties. The origin of
the unitary nature of this scattering from local defects has been ascribed to 
strong electronic correlations. While this issue is unsettled, it has in fact 
been suggested that a $Zn$ impurity may not simply act as a local moment 
vacancy because it induces a magnetic moment in its surrounding in the $CuO_2$ 
planes \cite{Xiao}. 

Transport data on $Zn$ doped metallic $YBCO$ have furthermore been used to 
infer a spatially extended nature of the effective impurity potential. In 
particular, a scattering cross section diameter larger than the $Cu$--$Cu$ 
distance has been deduced from the residual resistivity \cite{Ong}. 
Theoretically, for a model calculation of static vacancies in a Heisenberg 
antiferromagnet it has been shown early on that the ordered staggered magnetic 
moment is enhanced within a few lattice spacings around the vacancy 
\cite{Bulut}. Therefore, even in the hole doped system when spin correlations 
are dynamic and have a finite range the spatially extended structure of 
impurity effects is expected to persist. Electronic correlation effects thus 
provide a natural origin for the transport data. 

A series of model studies has been performed in recent years on the properties 
of dirty $d$--wave superconductors. Most of the analytical studies have assumed
$s$--wave impurity scattering while numerical simulations of impurities in 
Hubbard or $t$--$J$ type models have in fact demonstrated the expected 
effective, dynamically generated finite impurity potential range \cite{Hanke}. 
As a consequence higher order partial wave scattering channels become important
and lead to resonances in each channel. Impurity models for 
cuprate superconductors thus need to take into account the extended nature of 
the scattering centers \cite{Sascha,Wheatley}. 

\begin{figure}
\hskip2.cm
\psfig{file=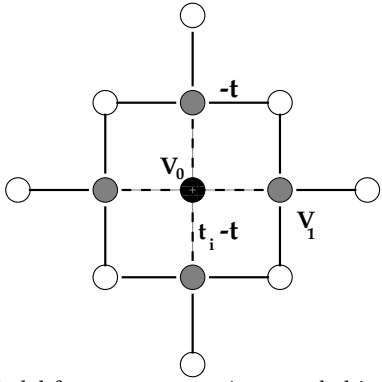,height=5.cm,width=5.cm,angle=270}
\caption[]{
Model for a nonmagnetic extended impurity on a square lattice. $V_0$ is the 
potential at the impurity center and $V_1$ the potential at the neighboring 
(nn) sites; $t_i$ is the change of the nn hopping amplitude of the host value 
$-t$.}
\label{fig1}
\end{figure} 

In this letter we explore the $T$--matrix in a minimal, physically transparent 
extended impurity potential model in a $d$--wave superconductor. In a partial 
wave decomposition the single impurity problem is solved exactly. Subgap 
resonances in the local DOS are strongly enhanced by a finite range of the 
potential and may thus be the origin for the unitarity limit scattering which 
was favoured in previous dirty $d$--wave studies for cuprate superconductors. 
This enhancement of the resonances improves the possibility for the detection 
of their highly anisotropic spatial structure in STM spectroscopy. Also the 
impurity induced zero energy DOS is found to be significantly increased for 
finite impurity concentrations as compared to the $s$--wave scattering limit 
for a local impurity potential.  

Specifically we model the scattering from impurity centers located at the sites
$\{{\bf l}\}$ on a square lattice
\begin{equation}
H_i=\sum_{\{{\bf l}\},\sigma,{\mbox{\boldmath$\delta$}}}\left[{V_0\over 4}n_{{
\bf l},\sigma}+t_i\left(c^+_{{\bf l},\sigma}c_{{\bf l}+{\mbox{\boldmath$\delta$
}},\sigma}+h.c.\right)+V_1n_{{\bf l}+{\mbox{\boldmath$\delta$}},\sigma}\right] 
\label{himp}
\end{equation}
(see Fig.1). Here, $\mbox{\boldmath$\delta$}$ connects to the nearest neighbor 
sites, $V_0$ and $V_1<V_0$ are potential strengths at the impurity center and 
its nearest neighbor sites (nn), respectively, and $t_i$ is an additional 
contribution to the nn hopping amplitude which we assume to be of opposite sign
to the pure host value $-t$. By focussing on a two parameter model for the 
specific parameter relation $V_1=\alpha^2 V_0/4$ and $t_i=\alpha V_0/4$ in 
Eq.(\ref{himp}) Fourier transformation leads to
\begin{equation}
H_i={V_0\over 4N}\sum_{{\bf l},\sigma}\sum_{{\bf k,q},{\mbox{\boldmath$\delta$}
}}e^{{\rm i}({\bf k-q})\cdot{\bf l}}V^\ast_{{\bf k},{\mbox{\boldmath$\delta$}}}
V_{{\bf q},{\mbox{\boldmath$\delta$}}}c^+_{{\bf k},\sigma}c_{{\bf q},\sigma}
\label{himpk}
\end{equation}
where $V_{{\bf k},{\mbox{\boldmath$\delta$}}}=1+\alpha e^{{\rm i}{\bf k}\cdot{
\mbox{\boldmath$\delta$}}}$. The virtue of the parameter choice becomes 
apparent in the factorization of the scattering potential which allows for an
algebraic solution of the single impurity $T$--matrix. $\alpha$ is the control 
parameter for the extension of the potential to the nn sites. The single 
particle Green function in the presence of the impurity is ${\hat G}({\bf k,k'}
,\omega)=\delta_{\bf k,k'}{\hat G}^0({\bf k,k'},\omega)+{\hat G}^0({\bf k},
\omega){\hat T}_{\bf k,k'}(\omega){\hat G}^0({\bf k'},\omega)$. All quantities 
are 2x2 matrices in Nambu particle--hole spinor space. The clean Green function
is given by $[{\hat G}^0({\bf k},\omega)]^{-1}=\omega{\hat\tau}_0-\Delta_{\bf k
}{\hat\tau}_1-\xi_{\bf k}{\hat\tau}_3=[\sum_{i=0,1,3}{\hat\tau}_ig_i]^{-1}$, 
where ${\hat\tau}_i$ ($i=1,2,3$) are the Pauli matrices, ${\hat\tau}_0$ is the 
unit matrix, and $\xi_{\bf k}$ is the quasiparticle tight binding energy 
relative to the chemical potential. Specifically, we consider a superconductor 
with a $d_{x^2-y^2}$ gap function $\Delta_{\bf k}=\Delta(\cos{k_x}-
\cos{k_y})/2$.

In order to explore local properties in the vicinity of an impurity we consider
first the problem of a single impurity located at the origin ${\bf l=0}$. The 
momentum dependent $T$--matrix is obtained from the set of equations:
\begin{eqnarray}
{\hat T}_{\bf k,k'}(\omega)&=&{V_0\over 4}\sum_{{\mbox{\boldmath$\delta,\delta'
$}}}V_{{\bf k}{\mbox{\boldmath$\delta$}}}^\ast{\hat t}_{{\mbox{\boldmath$\delta
,\delta'$}}}V_{{\bf k'}{\mbox{\boldmath$\delta'$}}} \\
{\hat{\underline t}}&=&\left[{\hat\tau}_0{\underline 1}-{\hat\tau}_3{\hat{
\underline t}}^0\right]^{-1}{\hat\tau}_3 \\
{\hat t}^0_{{\mbox{\boldmath$\delta,\delta'$}}}&=&{V_0\over 4}\sum_{\bf k}V_{{
\bf k}{\mbox{\boldmath$\delta$}}}{\hat G}^0({\bf k},\omega)V^\ast_{{\bf k}{
\mbox{\boldmath$\delta'$}}} \, ,
\label{tmateq}
\end{eqnarray}
where underlined quantities are 4x4 matrices with respect to the nn site 
coordinates next to the impurity center; note that their matrix entries contain
noncommuting Pauli matrices and therefore the matrix inversion in Eq.(4) is 
nontrivial. 

Using the square lattice basis functions $\gamma_{\bf k}^{(s/d)}={1\over 2}
\left(\cos{k_x}\pm\cos{k_y}\right)$ and $\gamma_{\bf k}^{(p1/p2)}={1\over 2}
\left(\sin{k_x}\pm\sin{k_y}\right)$ and thus decomposing into $s$--, $p$--, and
$d$--wave scattering channels we obtain the algebraic result for the 
$T$--matrix in the form
\begin{eqnarray}
{\hat T}_{\bf k,k'}(\omega)&=&V_0{\bf U}_{\bf k}^\ast{\hat{\underline D}}(
\omega){\bf U}_{\bf k'}{\hat\tau}_3\hskip0.2cm ,\nonumber \\
{\bf U}_{\bf k}&=&\left[\left(1+\alpha\gamma_{\bf k}^s\right),{\rm i}\alpha
\gamma_{\bf k}^{p1},{\rm i}\alpha\gamma_{\bf k}^{p2},\alpha\gamma_{\bf k}^d
\right]
\label{tsas}
\end{eqnarray}
where the matrix ${\hat{\underline D}}(\omega)$ is given by
\begin{equation}
{\hat{\underline D}}(\omega)=\left[\matrix{{\hat\tau}_1{\hat d}{\hat\tau}_1{
\hat S}^{-1}&0&0&{\hat a}^+{\hat D}^{-1}\cr 0&{\hat\tau}_1{\hat p}{\hat\tau}_1{
\hat P}^{-1}&{\hat a}^-{\hat P}^{-1}&0\cr 0&{\hat a}^-{\hat P}^{-1}&{\hat\tau
}_1{\hat p}{\hat\tau}_1{\hat P}^{-1}&0\cr {\hat a}^+{\hat S}^{-1}&0&0&{\hat\tau
}_1{\hat s}{\hat\tau}_1{\hat D}^{-1}\cr}\right] \, .
\label{amat}
\end{equation}
The structure of the matrix ${\hat{\underline D}}$ apparently implies a mixing 
of the $s$-- and $d$-- and the two $p$--wave scattering channels, respectively.
The matrix elements in Eq.(7) are determined by the 2x2 matrix functions
\begin{eqnarray}
\left({\hat s},{\hat p},{\hat d}\right)&=&{\hat\tau}_0-{V_0\over 4N}\sum_{\bf k
}\left(s_{\bf k},p_{\bf k},d_{\bf k}\right){\hat\tau}_3\left({\hat\tau}_0g_0+{
\hat\tau}_3g_3\right)\, , \\
{\hat a}^\pm&=&{V_0\over 4N}\sum_{\bf k}a^\pm_{\bf k}{\hat\tau}_3g_1{\hat\tau
}_1 \, ,\\
{\hat S}&=&{\hat s}{\hat\tau}_1{\hat d}{\hat\tau}_1-({\hat a}^+)^2
\hskip0.2cm ,\hskip0.2cm{\hat P}={\hat p}{\hat\tau}_1{\hat p}{\hat\tau}_1-
({\hat a}^-)^2 \, ,\\
{\hat D}&=&{\hat d}{\hat\tau}_1{\hat s}{\hat\tau}_1-({\hat a}^+)^2
\label{hatcoeff}
\end{eqnarray}
with the momentum dependent coefficients

\begin{figure}
\psfig{file=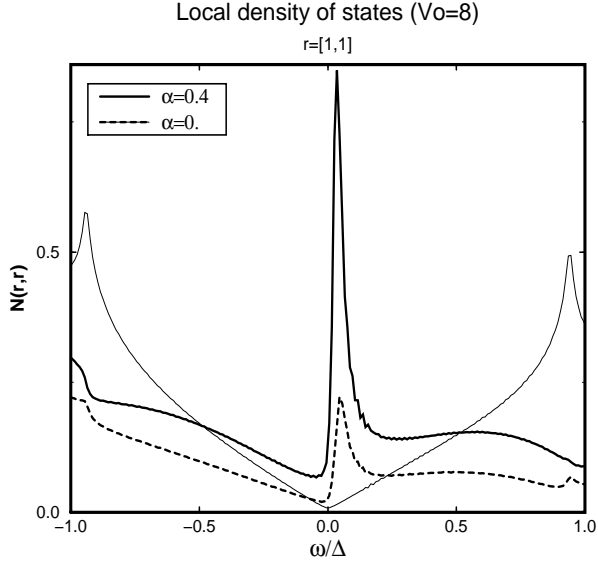,height=8.cm,width=9.2cm,angle=270}
\caption[]{
Frequency dependence of the local DOS at ${\bf r}=$ $(1,1)$ at $V_0=8t$ for an 
extended ($\alpha=0.4$, bold solid line) and a local ($\alpha=0$, dashed line) 
impurity in a $d_{x^2-y^2}$ superconductor. The impurity center is at the 
origin ${\bf r}=(0,0)$. The thin solid line shows the pure DOS. }
\label{fig2}
\end{figure} 

\begin{eqnarray}
s_{\bf k}&=&4+2\alpha^2+8\alpha\gamma_{\bf k}^s+4\alpha^2\cos{k_x}\,\cos{k_y}-
p_{\bf k} \, ,\nonumber \\
d_{\bf k}&=&2\alpha^2-8\alpha^2\cos{k_x}\,\cos{k_y}-p_{\bf k} \, , \nonumber \\
p_{\bf k}&=&\alpha^2-{\alpha^2\over 2}\left(\cos{2k_x}+\cos{2k_y}\right) \, ,
\nonumber \\
a_{\bf k}^\pm&=&2\alpha\gamma_{\bf k}^d\pm\left[2\alpha\gamma_{\bf k}^d+{
\alpha^2\over 2}\left(\cos{2k_x}-\cos{2k_y}\right)\right] \, .
\label{kcoeffs}
\end{eqnarray}
In the limit $\alpha\rightarrow 0$ the $T$--matrix becomes momentum independent
and reduces to the known $s$--wave scattering result of a local impurity 
potential \cite{Einzel}. Note, however, that for finite $\alpha$ the 
$T$--matrix ${\hat T}=\sum_{i=0}^3 T_i{\hat\tau}_i$ has components in ${\hat
\tau}_0$ and all three Pauli matrices while $T_1=T_2=0$ for $\alpha=0$. Given 
the explicit algebraic solution of ${\hat T}_{\bf k,k'}(\omega)$ the local DOS 
then follows from 
\begin{equation}
N({\bf r},{\bf r},\omega)={-1\over\pi N^2}{\rm Im}\,\sum_{\bf k,k'}e^{-{\rm i}{
\bf k}\cdot{\bf r}}{\hat G}_{11}({\bf k,k'},\omega)e^{{\rm i}{\bf k'}\cdot{\bf 
r}}
\label{dos}
\end{equation}
with $\omega+{\rm i}0^+$ implied.

We have evaluated the local DOS using a free tight binding band $\xi_{\bf k}=-4
t\gamma_{\bf k}^s+4t'\cos{k_x}\,\cos{k_y} -\mu$, with $t=1$ as the energy unit 
and $t'=0.4$, for a band filling $n=0.86$ and a $d$--wave gap amplitude 
$\Delta=0.3$. Yet, the conclusions are insensitive to changes in this parameter
choice. The potential at the impurity center is chosen as twice the bandwidth 
$V_0=8$. All results for the extended potential, i.e. finite $\alpha$, are 
compared to the local potential case $\alpha=0$.
 
In Fig.2 we have plotted the frequency dependence of the local DOS at a next 
nearest neighbor (nnn) site of the impurity center showing a sharp resonance at
a low subgap frequency $\omega_{res}$. The $p$--wave scattering channels are 
too weak to contribute an additional structure and the single resonance has a 
mixed origin in the $s$-- and $d$--wave channels. Quite remarkably, at this nnn
site the strength of the resonance is enhanced by a factor 4 when $\alpha$ is 
increased from $0$ to $0.4$. Comparably weaker resonances at $\pm\omega_{res}$ 
with weaker $\alpha$ induced changes appear along the $(1,0)$ and $(0,1)$
direction and this anisotropy directly reflects the nodal structure of the 
$d_{x^2-y^2}$ gap function \cite{Byers,Choi,Salkola}. 

In Fig.3 we show the oscillatory distance dependence of the local DOS along the
diagonal (and, in the inset, horizontal) direction at the resonance frequency 
$\omega=\omega_{res}$. The enhancement due to a finite range of the potential 
is largest near the impurity center where the resonance is strongest. In 
comparing the results for the diagonal and the horizontal direction we 
recognize that the direction dependence of the resonant structure is amplified 
by the finite potential range thereby improving the possibility for detecting 
the anisotropic local DOS structure by STM in the vicinity of the impurity.  

\begin{figure}
\psfig{file=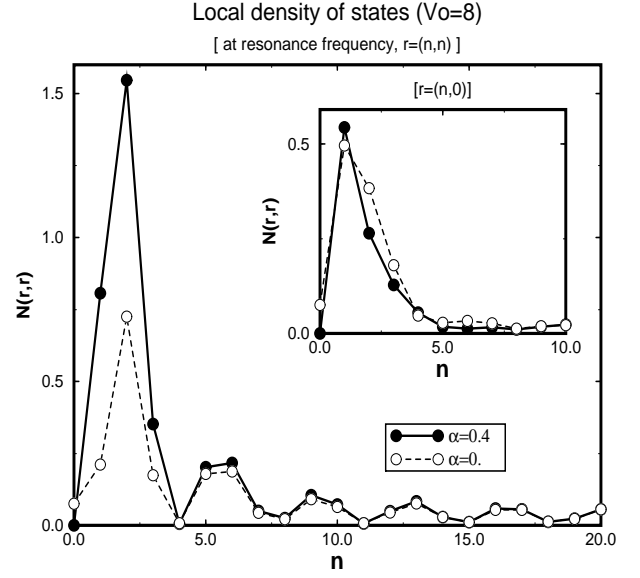,height=8.cm,width=8.8cm,angle=270}
\caption[]{
Distance dependence of the local DOS at the resonance frequency $\omega=+
\omega_{res}$ along the diagonal (main figure) and horizontal (insert) 
direction away from the impurity center at ${\bf r}=(0,0)$.}
\label{fig3}
\end{figure} 

While this is already a very promising result for local experimental techniques
we extend the algebraic results for the single extended impurity to a finite 
impurity concentration $n_i$ within the self--consistent $T$--matrix (SCT)
approximation. In the SCT approximation we obtain the impurity position 
averaged self--energy as ${\hat\Sigma}({\bf k},\omega)=n_i{\hat T}_{\bf k,k}(
\omega)$. With ${\hat G}({\bf k,k},\omega)=[({\hat G}_0({\bf k},\omega))^{-1}-
{\hat\Sigma}({\bf k},\omega)]^{-1}$ replacing ${\hat G}_0$ in the above exact 
formulas for the single impurity $T$--matrix we solve the resulting 
self--consistency equations for ${\hat G}$ by iteration without the usual 
assumption of particle--hole symmetry. Note that ${\hat\Sigma}=\sum_{i=0,1,3}
\Sigma_i{\hat\tau}_i$, i.e. the $\tau_1$ component of the self--energy is 
finite for $\alpha>0$. 

Fig.4 shows the expected result for the SCT DOS for a low impurity 
concentration $n_i=1\%$. At first sight the zero frequency DOS increase with 
turning on $\alpha$ might seem weak. However, within the chosen parametrization
$\alpha=0.5$ translates into $V_1/V_0=1/16$; i.e. if the nn impurity potential 
strength is only $6\%$ of the impurity center value and accompanied by a 
suppression of the hopping amplitude to the impurity center, $N(\omega=0)$ 
increases by $\sim 15\%$ which is indeed significant. 

In summary, we have solved a physically motivated extended impurity potential
model in a $d_{x^2-y^2}$ superconductor. The finite range of the potential is
argued to be dynamically generated around a local impurity from strong 
electronic correlations in the host system and thus relevant for high--$T_c$ 
superconducting materials. While the model itself describes a static potential 
with a selected parameter relation it has the advantage to allow for an 
algebraic solution by decomposition into $s$--, $p$--, and $d$--wave scattering
channels. The significant enhancements of the impurity induced resonances by 
the finite potential range clearly improve the proposed possibility for probing
the highly anisotropic local DOS structure near the impurity by STM experiments
and may thus serve as an alternative fingerprint of the gap symmetry in cuprate
and other unconventional superconductors. Furthermore, due to the physical 
relevance for the description of impurity effects in high--$T_c$ 
superconductors some of the previous model studies with $s$--wave scattering
from local impurity potentials may be reconsidered for the effects of an 
extended potential range for which the presented model can serve as an 
algebraically tractable tool.

\begin{figure}
\psfig{file=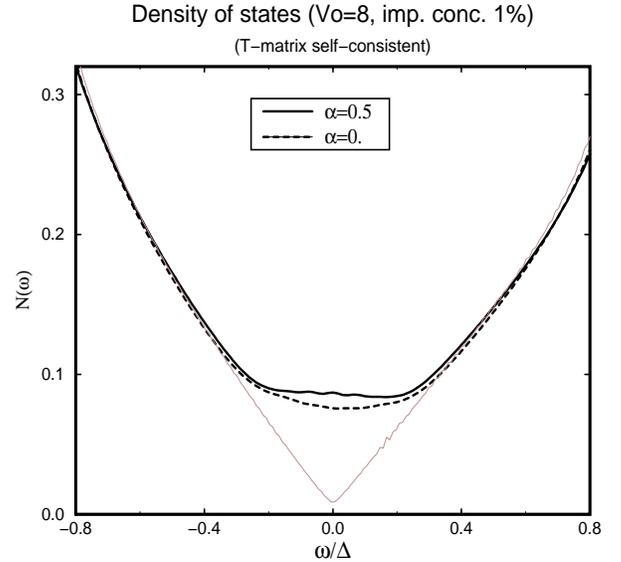,height=8.cm,width=9.2cm,angle=270}
\caption[]{
Impurity position averaged DOS from the self-- consistent $T$--matrix 
approximation for an impurity concentration of $n_i=1\%$. For the meaning of 
the different lines see Fig.2.}
\label{fig4}
\end{figure} 

A.P.K. acknowledges support through a Heisenberg fellowship of the Deutsche 
Forschungsgemeinschaft (DFG). A.P.K.'s research was performed within the 
program of the Sonderforschungsbereich 341 supported by the DFG. T.P.D. would
like to acknowledge helpful discussions with Drs. J. Byers and M.E. Flatt\'e.

\end{document}